# Perception, Prestige and PageRank

David Zeitlyn[1] and Daniel W. Hook[2,3,4]

[1]Institute of Social and Cultural Anthropology, School of Anthropology and Museum Ethnography, University of Oxford, 51 Banbury Road, Oxford, OX2 6PF, UK
[2] Digital Science, 4 Crinan Street, London, N1 9XW, UK
[3]Centre for Complexity Science, Imperial College London, London SW7 2AZ, UK
[4]Department of Physics, Washington University in St Louis, St Louis, MO, USA

Abstract

Academic esteem is difficult to quantify in objective terms.  Network theory offers the opportunity to use a mathematical formalism to model both the esteem associated with an academic and the relationships between academic colleagues.  Early attempts using this line of reasoning have focused on intellectual genealogy as constituted by supervisor student networks. The process of examination is critical in many areas of study but has not played a part in existing models.  A network theoretical "social" model is proposed as a tool to explore and understand the dynamics of esteem in the academic hierarchy.  It is observed that such a model naturally gives rise to the idea that the esteem associated with a node in the graph (the esteem of an individual academic) can be viewed as a dynamic quantity that evolves with time based on both local and non-local changes in the properties in the network. The toy model studied here includes both supervisor-student and examiner-student relationships. This gives an insight into some of the key features of academic genealogies and naturally leads to a proposed model for "esteem propagation" on academic networks.  This propagation is not solely directed forward in time (from teacher to progeny) but sometimes also flows in the other direction. As collaborators do well, this reflects well on those with whom they choose to collaborate and those that taught them. Furthermore, esteem as a quantity continues to be dynamic even after the end of a relationship or career.  In other words, esteem can be thought of as flowing both forward and backward in time.

*Keywords*:  Research Information Management, Bibliometrics, Complex Network, Ranking, Kinship

## Introduction

The use of network theory to represent and study the nature of academic relationships is not new. Co-authorship and citation graphs are used, for better or for worse, to meet the data requirements of academic appointments panels, tenure review committees, funder processes and government assessment activities around the world. However, outside the context of research evaluation, there remain some practical and important uses of graph theory: To understand the semantic proximity of research fields; to observe and predict the emergence of new fields; to quantify the connectedness of individual researchers; and, to locate researchers who form pivotal relationships and bridge different fields with their research.

The idea of an intellectual genealogy as a set of relationships between supervisors and research students is well established and leads naturally to the concept of an intellectual lineage [11,16]. While the supervisor-supervisee relationship is the most commonly recorded and studied relationship it is not the only link that can be used to formulate an intellectual genealogy. Based on an empirical case study (used as a toy model), this paper proposes a framework to quantify esteem flow and ranking on an academic network. Researchers habitually try to contextualise colleagues by their relationships. It is generally accepted that in the early stages of a career whether someone's supervisor was "good", whether their examiner was "well-known" and whether their research collaborators are "recognised" are important factors in later success. Of course, "good", "well known" and "recognised" are all subjectively defined but it is clear that the perception generated by being "chosen" by accepted prestigious individuals is an important one, not just in academia. Such observations have recently become a "hot topic" as analysis by Ma et al. [30] suggests that women benefit less from this ecosystem than their male counterparts.

We conjecture that a deeper analysis of the network approach that we have suggested here would highlight further examples of women who have been overlooked but who have had significant and unsung influence on a field.

Over the last 50 years' doctorates have become established as a gateway to a career in the academy. As a result, doctorates have become a common currency in academia and are now often a rapid proxy to locate a researcher in some mental map of the research landscape.  That is to say that especially at the beginning of an academic career, before considering the number of publications, books or scholarly outputs created; the counts of citations, attendance numbers at public viewings or critical reviews of performances; commonly asked questions include: Where did the researcher in question study? In what department? And, With whom? The answers to these questions instantly give a picture of the researcher in the internal ranking system of the questioner.  If the institution is well known, the department of note or the supervisor famous, then an expectation is already set by this context.  This impression is not easy to make tangible since it is a highly-subjective measure of academic prowess and is personal to each evaluator and their own academic collaborations, interactions and experiences. Hence, while many decisions are influenced by this type of esteem measure, it is not easy to define and is impossible defend in objective terms.

In this paper, we carry out an initial exploration of a dataset that we have created by bringing together data on one subject area from several sources and formulate some conjectures based on these data and an initial analysis.  In particular, we look beyond the supervisor-supervisee doctoral thesis graph. We consider connections mediated by the doctoral thesis but instead of considering only the supervisory relationship we also consider the graph formed by taking the examiners (so covering all parents) of a doctoral thesis and the candidate (child).

There is a rich and complex sociology behind thesis examination that needs to be understood before we can extend the supervisor-supervisee network model with confidence. Our initial approach was to consider all dissertations (masters level, and doctoral level) and weight the relationships associated with the level of the qualification, but this adds significant additional complexity: It is unclear what would constitute a fair weighting, it is unclear whether the a masters dissertation included novel research or was simply a literature review and, finally, the number of masters students who do not carry on to academia is high and hence many of these nodes remain unconnected and merely add data complexity without adding content. Additionally, in many universities, a student "failing" a doctoral qualification may be given an MPhil or other Master-level qualification. It is not easy to tell from the records that are kept publicly which course was originally intended. Having a number of failed students may be an interesting signal to study but could be symptomatic of many different factors and hence becomes quickly too complex (not to mention controversial) to access as a problem in an initial study. As a result of these complicating factors, we have focussed on successful doctoral theses. This simplifies some of the data issues and focuses the study on a richer network where the participants are more likely to remain part of the research ecosystem and hence contribute to the graph more actively by becoming examiners themselves. Another weighting effect that should be considered is the relative importance of examining versus supervising from field to field. Although the supervisor spends more time with the student during the performance of their work, in some fields, the examiner of the thesis will be rated at a similar level of importance (especially where there are doctoral committees).  Since we are dealing with modelling perception in this case, the relative importance between supervisor and examiner may even be examiner (and supervisor) dependent.  In the case that an extremely eminent individual examines the thesis, this may have more bearing than a relatively less known supervisor.

Further complexities still exist even with our initial simplification: In some universities, defences are public and there is a team of examiners, in other institutions, the viva is closed and attendance is limited to the candidate and two examiners.  In some areas it is acceptable for a candidate's supervisor to attend the viva, in others the supervisor may be one of the examiners, in yet others the supervisor may not be allowed to attend.  Some universities require that one or more of the examiners must be from outside the institution (so-called 'externals').  Many institutions require that there is no formal relationship between the candidate and the examiner (i.e. they may not have written a paper together prior to the examination).

Even beyond these conventions there is an intricate art to the choice of examiners.  In some fields the external examiner is viewed as being almost as important as the supervisor since they act as the gatekeeper who allows a candidate to pass from student to a member of the academy.  Passing a defence with a well-known, respected or highly-established academic can be a mark of esteem and can be helpful for a young researcher.  Of course, choosing a very well-known academic as a panellist can be risky since failing a viva can be quickly reported around the academic social network and can reflect badly on the student, supervisor, and even the department or university.  Often, panellists are co-authors of one of the supervisors.  All these factors contribute to a subtle and interesting network for study.

There are also other important effects to bear in mind when considering the network of panellists that relate to institutional esteem and to practicality: it may be that there is a systemic internal bias for institutions to choose examiners from universities that are currently deemed to have higher status than the source institution; for practical purposes geography can play a significant role in that it is easier and cheaper to get an external examiner from a nearby university than from further afield where logistical issues makes the arrangement more difficult or more expensive.  Of course, the factors of status and geography interfere in a complex fashion.  Furthermore, the choice of possible externals is also influenced by judgements by supervisors about the merits of individual students.  A chosen examiner must be available, must know the topic concerned and be appropriate both in terms of prestige / status and the supervisor's judgement of the merit of the thesis.  It is plausible that across the academy the effects of these merit judgements cancel out leaving distance and status as the most significant factors.

A number of data sets suggest themselves as a starting point for studying this area: there are several online academic genealogy projects holding PhD student-supervisor data [19,29]: We collated data from the following online collaborative sources: Academic tree http://academictree.org/; Primatology Tree http://www.physanthphylogeny.org/ and PhD Tree http://phdtree.org/.  Goydar worked on the comparative prestige of sociology departments based not on publications but on the prestige of the universities awarding the PhDs of staff members [10].  The latter approach seems to capture the entrenched position of Oxford, Cambridge and few other Russell Group Universities in the UK together with the Ivy League and similar high-profile institutions in the US.  It is interesting to note that while the identity of an institution typically won't change in the short term, its level of esteem or prestige may, although typically within the lifetime of a single academic career the esteem of an institution can mostly be considered to be stable. There may be more, faster variation in the esteem of individual departments [17].

This paper is organised as follows: In Sec. 1 we explain how we created and curated the case study dataset that we then go on to use as a toy model to explore the ideas laid out above. In Sec. 2 we summarise the analysis of the dataset. In Sec. 3 we discuss a potential model to explore esteem propagation across an academic network. In Sec. 4 we include a brief discussion of the research and suggest some directions for further research.

## 1. Primatology as a toy model

For our study, we have chosen to study a dataset in the field of primatology and have created a network as described below. We are interested in a genealogical or kinship approach and so each node in our network corresponds to a person. Each examination defines a relationship between a set of nodes or individuals. The role that each person plays in each exchange is not a property of the node but rather of the link (the degree/examination). Hence, there are two types of directed edge in our network. Each edge corresponds to a single instance of a degree / examination and hence is associated with a single student who is undergoing an examination. There are two types of edge: supervisor-student edges and examiner-student edges and so a single examination in codified in a number of different edges. We have not attempted to codify other types of relationship in the graph such as examiner-examiner, supervisor-examiner or supervisor-supervisor edges – such relationships are left to be inferred from the graph through the examination events.

There is an implicit hierarchy in the network since there are dates associated with the examination events and the assumption (in general) is that the least experienced researcher will be the one undergoing examination or supervision. As a result, this is a temporal (edge) network graph of the type discussed by Holme & Saramaki [31] or Masuda and Lambiotte [32], although we have not exploited this in our work to date. Figure 1 shows a simple instance of a 3-generation academic genealogy in which the student E in Generation 2 is the supervisor of G in the 3rd generation and who draws on her previous supervisors as examiners and is co-supervising the student with one of her examiners.

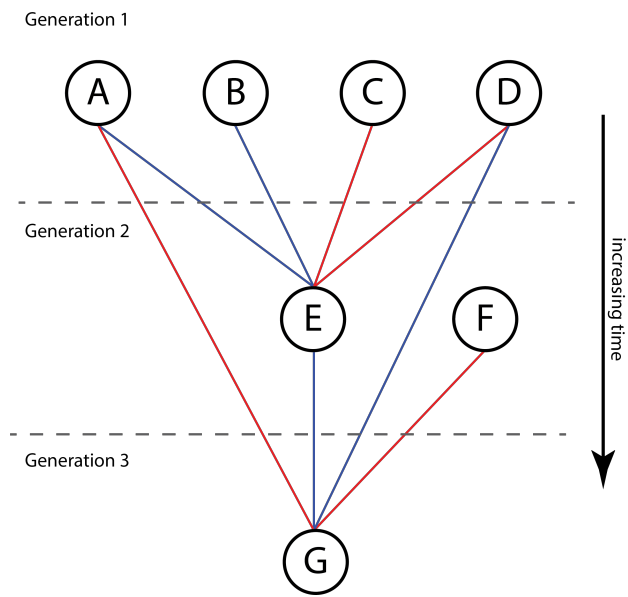

*Figure 1: Simple model for academic hierarchy involving both supervisor-student links and examiner-examinee links. Red lines represent examination relationships and blue lines represent supervisor/advisor relationships. In this diagram, A and B supervise E, who is examined by C and D; E then supervises G with D, who is in turn examined by A and F.*

In practical terms, finding data to formulate this graph is challenging and many value judgements must be made to create something that can sensibly represent an area. As we were looking for a graph containing links based on examinations, we needed to find a field where data was sufficiently readily available and the intellectual genealogy work of Sussman and Kelley was generously shared in order to provide a starting point for our own research [7]. The database of PhD supervisor-supervisee relationships provided by Sussman and Kelley was heavily supplemented by adding in doctoral examiner and committee data for the first time.

Beyond data availability, there are several reasons that we chose primatology for our toy model. Primatology is an interdisciplinary subject spanning observational field studies via captive studies to genetics. Primatologists are to be found in Zoology, Psychology, Biology, Medicine, Archaeology and Anthropology departments. This means that we are able to access a number of different subject styles to get a more universal picture. From a practical perspective, the size of the field of primatology is sufficient to be interesting but small enough that we are able to handle the data for cleaning purposes as well as for the purposes of drawing meaningful conclusions.

Kelley and Sussman's original dataset concentrated on American field primatologists and identifies the supervisory linage of Sherwood Washburn as the defining characteristic of the discipline. In the current work, we cast our net wider restricting ourselves to neither a methodological orientation or a sub-continent. Nevertheless, Washburn's lineage emerges as central. From a technical perspective, we have set further limits on what constitutes primatology for our purposes – specifically, we have not included theses on humans. Decisions on the inclusion or exclusion of theses were chosen by Zeitlyn on a case by case basis. It was not possible to automate or map this process to an algorithm. Classification of work in a narrow field remains a challenging problem. In order to do this in a reproducible manner using machine learning or classification algorithms, it is necessary to have a sufficiently large body of pre-tagged data, this is something this is not readily available in the area of primatology (in fact the current dataset associated with this paper may be the first to provide a learning set on which to base algorithmic approaches). We also need to specify what types of thesis are appropriate to include in our study: The theses that we deemed relevant for analysis had to meet the following two criteria: Firstly, only doctoral theses that resulted in an awarded qualification were included (this means that masters-level dissertations, even if related to research results were not considered); secondly, only theses related to non-human primatology were included. The core dataset was derived by searching the Proquest Dissertations & Theses database using the search string:

*subject("Physical anthropology") and (angwantibo or aye-aye or bonobo or bushbaby or galago or guenon or guereza or malbrouck or mandrill or muriqui or nycticebus or potto or talapoin or vervet) not (sifaka or loris or lemur or bushbaby or tarsier or marmoset or tamarin or capuchin or titi or saki or uakari or howler or macaque or mangabey or baboon or mangabey or drill or colobus or lutung or surili or douc or monkey or langur or gibbon or ape or orangutan or Gorilla or chimpanzee) or primatology)*

We include a number of exemplar cases here to illustrate the types of decision that have been made: Maki's 2013 'The Biomechanics Of Spear Throwing: an Analysis of the Effects of Anatomical Variation on Throwing Performance, with Implications for the Fossil Record' was deemed out of scope.

When considering the inclusion of supervisors and examiners we have taken the view that anyone supervising or examining a primatologist should be included (whatever their formal subject affiliation) since their supervision/examining has a bearing on primatology. Conversely, not all those supervised or examined by a primatologist may be primatologists so not all the students of a given parent (even of Washburn) have been included. This is a subject based project.  The dataset that we compiled and studied is available, together with an interactive visualisation [25].

It is important to note that not all individuals considered to be primatologists are included in our network for the simple reason that their doctoral studies were not on primatology. Hence the doctorates of several prominent individuals who became primatologists only after completing a doctorate on other subjects are not included. Examples include Clarence Raymond Carpenter since his 1932 Stanford Ph.D. was in psychology and ornithology even though Kelley and Sussman describe him as 'the first true field primatologist' on the basis of his post-doctoral work. He is, however, included in our database as supervisor and examiner. Another similar case is that of Thelma Rowell.

Among other examples we note Dr Klára Petrželková, whose 2003 PhD from Masaryk University in the Czech Republic was on bats and so deemed out of scope. It was only subsequently that she started working on primates. Petrželková does, however, appear as a supervisor and examiner of primatologists. Other notable examples include James R. Roney, 'Psychological and hormonal responses of men to sensory cues from women' (2002) https://catalog.lib.uchicago.edu/vufind/Record/4733089; Harry Israel's Stanford PhD on albino rats (1930). The last has a further significance since the author changed his name and subsequently was better known as Harry F. Harlow. Another, more recent case of male name changes is Adam Clark Arcadi who writes on his web page http://anthropology.cornell.edu/people/faculty-list/adam-clark-arcadi.cfm 'Prior to 1996 I published under the name Adam P. Clark', last accessed 1 Oct 2015. We have several cases of women changing name on marriage and / or divorce. This leads to challenges in carrying out person disambiguation on the dataset. While analysis has been carried out to use standard pattern matching approaches to deduplicate individuals in the data as well as some manual work, only in cases where we explicitly know about changes of name have we been able to take this into account in the underlying data.

Since in some systems of doctoral assessment supervisors are also examiners there was some duplication in the data. We left this data in the main dataset set since it may be of interest (it represents a meeting between the groups of co-examiners, the juries), but in the network analysis reported here we omitted examiners who were also supervisors, on the basis that their role in the examination process was different from the other examiners. They were present as a function of their role as supervisor, they had not been asked solely to assess the submitted thesis.

We note this points to an ambiguity in some US doctoral committees, which are constituted from the start of the research and so combine supervisory and examining roles. We have taken the one or two committee members designated as 'primary supervisors' as such and treated the remainder of the committee as examiners.

In our data the maximum number of examinations for an individual is 56 and the maximum number of supervisions is 32 (not the same person). Furthermore, we note that some individuals have several doctorates (quite apart from honorary doctorates which we are not discussing at present). An extreme case is William Clement McGrew who has 3 doctorates at least 2 of which fit our criteria.

## 2. Analysis and results

As active primatologists Kelley and Sussman had a natural intuition that Washburn was a central figure in primatology and, indeed, were able to demonstrate the importance of his lineage in America. They also document some other lineages in other countries (e.g. that associated with Robert Hinde in Cambridge) and some others in America. In our work, we seek to remove the intuition component and show that network properties and statistics confirm intuition and can be used in place of intuition to identify key actors in a field.

In our analysis we used Pajek and the SNA package in R to calculate network statistics. Overall, we captured and codified 7557 examinations, 4628 supervisions related to a total of 8443 qualification "events", comprising 9828 connections between examiners and vivas and, 3486 connections between supervisors and students. Table 1 summaries some of the key network statistics.

*Table 1: Summary of key network statistics for primatology toy model including, number of disconnected components, size of largest connected component and network density.*

| Network | Number of nodes | Number of edges | Number of connected components | components containing more than 10 nodes | Nodes in largest comp | Percentage of network in largest comp | Network density |
|---|---|---|---|---|---|---|---|
| Examinations | 7557 | 9828 | 486 | 31 | 4939 | 65.36527 | 0.0001721626 |
| Supervisions | 4628 | 3487 | 1188 | 27 | 1119 | 24.18414 | 0.0001628629 |
| All | 8443 | 13314 | 534 | 32 | 5764 | 68.27766 | 0.0001868398 |

We take the genealogical metaphor as a guiding principle, especially since our data are bilateral, which means that we can explore the possibility of whether there are intellectual analogues to cousin marriage. Forms of asymmetrical marriage among distant kin are well testified from round the world and are an old staple of anthropological kinship theory. We note that this appears to make the somewhat recherché and old-school topic in Kinship Theory of preferential patrilateral cross-cousin marriage, as practiced among nomadic pastoralists to keep herds intact, relevant to the study of social networks. In network analysis this is equivalent to asking whether the network contains closed loops or cycles. That is to ask whether starting from an individual is it possible to find a path back to that individual in which linking people occur once only. Of course, since our data are time-ordered, it is simple to remove these links if needed. Kinship has been one of the central parts of academic anthropology for more than a century [23,26]. The study of kinship concerns both the tracing of descent (the ways in which people identify themselves to one another) and the arrangement of marriage, who is marriageable and who cannot be married.

Loops occur for a number of reasons.  For example, hypothetically two students of the same supervisor might in time come to have their own students. When looking for an examiner, Supervisor A may think of their former peer as being suitable since they come from the same disciplinary tradition.  This would create a very small cycle, see Figure 2.

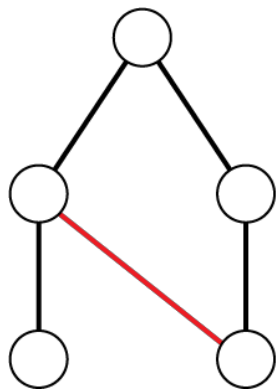

*Figure 2: Simple closed loop showing the relationship where one examiner had the same intellectual "father" or "mother" as the supervisor of the student being examined.  Black lines show supervisory relationships.  Red lines show examiner relationships.*

We note that it might seem wrong to "return the favour" as this may be termed 'incestuous' see Figure 3.  Note the naturalness of the genealogical metaphor. In a similar vein the tendency of universities to appoint their own graduands could also be loosely described as being incestuous but taking the kinship metaphor seriously would suggest this is more accurately described as a form of intellectual endogamy (see Godechot [4-6,13-15] whose title talks of 'inbreeding'). However, due to the availability of subject experts in a field or subfield it is the case, anecdotally, that this practice does go on.

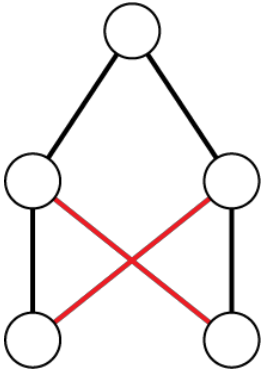

*Figure 3: Incestuous relationship loop, in which two supervisors who are children of the same intellectual parent, examine each other's students.  Black lines denote supervisory relationships.  Red lines denote examiner relationships*

If the reciprocity was delayed by a generation it might seem more appropriate: we could imagine either students of two students of one supervisor examining each other ('first cousin once removed' examining) or the return occurring in a later generation, see Figure 4.

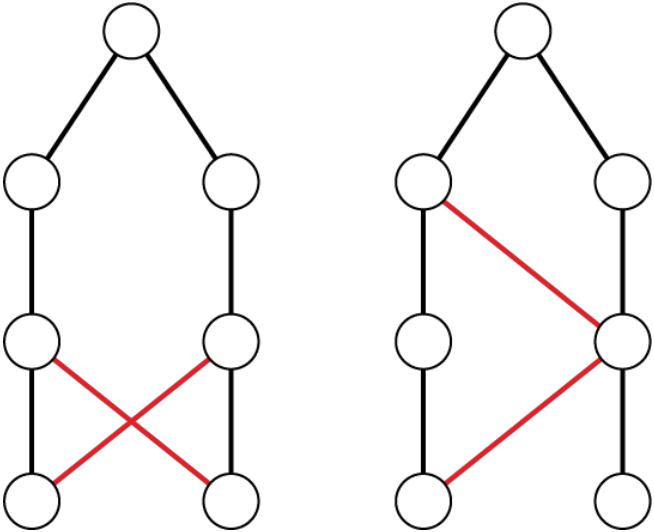

*Figure 4: Supervisor / examiner relationships delayed by one generation.  Black lines denote supervisor relationships. Red lines denote examiner relationships.*

We should stress that these sort of loops may be emergent properties of localised strategic decision making rather than being deliberately crafted. We are not suggesting that examiners are being chosen because of being intellectual uncles or aunts of the student concerned. In some cases, however, these are the people best placed to act as examiner and unintended consequences of such assessments is that cycles are created.  In our data, we found 48 examples of simple closed loops and 36 examples of Incestuous closed loops.

We have gone to a lot of trouble to set up a network to study.  However, there are some simple statistics to collect to get a sense of the graph before we look at network statistics.  Since we are particularly interested in esteem, a crude measure of esteem may be related to the number of doctoral students that a researcher has supervised.  Being successful, prominent or of high status in a discipline is likely to lead to an individual attracting research students and to be being asked to serve as examiner for many others. On the face of it there is potentially a gradient from someone with a newly minted doctorate who has neither themselves been an examiner nor supervised any students to a senior figure who at retirement has supervised and examined many students. To explore this intuition, we took the topmost individuals in the lists of most supervisees (Table 2) and most examinees and collated the awards and other tokens of esteem they had received.

*Table 2: Top supervisors in the field of primatology by number of supervisees.  Only those with more than 10 students are shown.*

| (node_id) Name | Number supervised |
| --- | --- |
| (Merge_02898) Dunbar, Robin I. M. | 32 |
| (AU_16114) Sussman, Robert Wald | 29 |
| (AU_16017) Jolly, Clifford J. | 27 |
| (Merge_07191) Martin, Robert D. | 25 |
| (AU_16069) Washburn, Sherwood Larned | 23 |
| (AU_16040) Lee, Phyllis Chadwick | 21 |
| (TempID_211) Hinde, Robert | 21 |
| (AU_16019) Wrangham, Richard W. | 20 |
| (AU_16375) Dolhinow, Phyllis Jay | 20 |
| (Merge_09577) Rodman, Peter S. | 20 |
| (AU_16024) Richard, Alison Fettes | 19 |
| (AU_16048) McGrew, William C. | 17 |
| (AU_16091) Tuttle, Howard Russell | 17 |
| (AU_16105) Snowdon, Thomas Charles | 17 |
| (AU_16085) Simonds, Paul Emery | 16 |
| (TempID_537) Whiten, Andrew | 16 |
| (AU_13219) Wright, Patricia Chapple | 15 |
| (AU_16492) Nishida, Toshisada | 15 |
| (AU_16562) Van Schaik, Carel P. | 15 |
| (AU_16088) Devore, Irven | 14 |
| (AU_16096) Poirier, Eugene Frank | 14 |
| (AU_16190) Cords, Ann Marina | 14 |
| (AU_16097) Bramblett, Claud Allen | 13 |
| (Merge_04834) Holekamp, Kay | 13 |
| (AU_13024) Glander, Kenneth Earl | 12 |
| (AU_16018) Oates, John F. | 12 |
| (AU_13047) Godfrey, Laurie Ann Rohde | 11 |
| (AU_13151) Harrison, Terry | 11 |
| (AU_13175) Watts, David Peter | 11 |
| (AU_16106) Pilbeam, Roger David | 11 |
| (AU_16118) Delson, Eric | 11 |
| (AU_16125) Kay, Richard Frederick | 11 |
| (AU_16137) Fedigan, Linda Marie | 11 |
| (Merge_01276) Buikstra, Jani E. | 11 |

To motivate the use of a network-statistic approach to gain insights we look at a slightly more fine-grained statistic that is natural to derive from the course-grained statistics in Table 2. The number of examinations in which an examiner has participated, in Table 3, can be misleading since in the committee and jury systems many supervisors are members of their own students' doctoral committees/juries. We have attempted to allow for this by counting only examinations of students that are supervised by others and by weighting the figures to reflect the difference between internal and external membership of juries and committees. For the UK examination system where the norm is only two examiners we have included only the external examiners and have not recorded internal examining.

To reflect the relative perceived merit of internal versus external examination we introduce an arbitrary relative weighting to make a point. In our weighting of committee /jury members external members are weighted 1, internal at 0.5. Where affiliation is not recorded, as sometimes occurs in the data (in our dataset, 0.4% of thesis authors and 6% of authors and examiners overall lacked affiliation information), we assume they are internal so weight them by 0.5.

Note that if we look purely at the data in Table 3, this is a difficult half-way house between a very pure high-level statistic and a measure that attempts to give us context. If you look in isolation at the number of examinations then you may think that Clifford Jolly is the most highly esteemed academic.  However, number of external examinations might be a better proxy and so Jolly, Wrangham and Wright all do well on this measure.  However, using our arbitrary weighting Sussman scores extremely highly. The arbitrariness of the weighting that we've introduced makes sense instinctively but skews the numbers.  So, our attempt to account for effects that we know about by adding in an arbitrary variable only makes things worse.

Indeed, the problem is significantly more complex when you try to add further context. At the centre of the problem is the lack of context built into the statistics that we have available. In what communities do these individuals sit: are they at large institutions? Do they have funding to travel due to generous research funding environments? Are they based in one of the geographical regions in which there are many universities (for example New York, Boston, London or Melbourne) and so invitations to participate in external evaluations may be more plentiful since travel costs can be kept low?  No high-level analysis is ever going to capture that level of nuance and for a group of thousands of people it is impossible to capture enough variables to truly represent this.  However, we argue that the structure of the network that we've build should encode some of this "landscape" information and hence network statistics should provide more contextualised norms.  In some sense, the weighting that we codified manually should already be encoded in the network structure and should emerge through the ranking.

*Table 3: Ranking of primatologists by examining engagements.*

| (node_id[*]) Name | Number examined | Number own students examined | Number of others examined | Weighted no. examinations (External + 0.5 internal) |
|---|---|---|---|---|
| (AU_16114) Sussman, Robert W. | 41 | 15 | 26 | 33.5 |
| (AU_16017) Jolly, Clifford J. | 56 | 36 | 20 | 38 |
| (Merge_09577) Rodman, Peter S. | 22 | 6 | 16 | 19 |
| (AU_16375) Dolhinow, Phyllis J. | 18 | 3 | 15 | 16.5 |
| (AU_13219) Wright, Patricia C. | 35 | 21 | 14 | 24.5 |
| (AU_16096) Poirier, Eugene F. | 19 | 7 | 12 | 15.5 |
| (AU_16019) Wrangham, Richard W. | 35 | 24 | 11 | 23 |
| (AU_16097) Bramblett, Claud A. | 17 | 6 | 11 | 14 |
| (AU_16148) Fleagle, John G. | 45 | 35 | 10 | 27.5 |
| (AU_16091) Tuttle, Howard R. | 19 | 9 | 10 | 14.5 |
| (Merge_05008) Howell, Francis C. | 18 | 8 | 10 | 14 |
| (AU_16069) Washburn, Sherwood L. | 16 | 6 | 10 | 13 |
| (Merge_04834) Holekamp, Kay | 14 | 4 | 10 | 12 |

*node_id refers to the id column of the DATA_names.txt file in [25].

Although there was a high degree of correlation between many of the network and non-network measures there were some individuals identified by network measures who would have been missed by the simple counting exercise and which, perhaps, more closely represent an "insider's" intuitions e.g. that Washburn was an important figure, founding a lineage. Significantly, though Washburn is identified as important by all measures so is Howard Russell Tuttle.

Given that network statistics have a slightly different flavour to mere counting it is important to understand how closeness centrality and betweenness centrality work. In his recent work, *the Square and the Tower*, Naill Ferguson gives an excellent account of network statistics [24], while more technical treatments may be found in [27,28]. Closeness centrality is calculated as the sum of the length of the shortest paths between the node and all other nodes in the graph. It is important because it gives a sense of the directness connection of any node to any other node. In academic terms, those with high closeness centrality tend to be established academics with developed social networks arising from a rich collaborative career and significant conference attendance or positioning in a connected academic institution. In some sense, closeness centrality can be a good measure of "establishment" or "fame". The *betweenness centrality* of each node is defined in terms of the shortest paths between pairs of points in the graph. The node with the highest betweenness centrality is the node through which the highest number of shortest paths pass. In a sense, this is the node which is most likely to be "on the way" from any random choice of node to any other random choice of node. Betweenness gives us a measure of how strategically a node is connected. For example, if a graph may be decomposed into two equal internally-connected but overall disconnected graphs and one new node is

introduced with a single connection to each of these two graphs, then it is instantly a candidate to

be the node with the highest betweenness since all paths must pass through that node which wish

to travel from one graph to the other.  In academic terms, we find people with high betweenness

in many scenarios but particularly in the early stages of emergence of interdisciplinary

communities.  Those academics who connect two disparate communities have high betweenness

in co-authorship networks as there are few links between those communities.  Betweenness can

be a good measure of innovation or a signal for certain types of collaboration. We have

calculated these two values for all academics in our network and show the most highly ranked in

Table 4.

*Table 4: Lists the top ten individuals for five different network measures (and those
supervising/examining >10 theses).  We then removed those identified by only one measure then look at
the individuals who are prominent by several different measures. Clossness Centrality and Betweenness
centrality have the standard definitions. Output domain (degree) of an individual is the number of all
people to whom each actor is connected.  Output proximity prestige of an individual is the proportion of
all other vertices in their output domain divided by the average distance from this individual to the other*

*people in their output domain (that is, normalized output domain / average distance). The citation weight of an individual (using the Search Path Count algorithm) is the normalised sum of direct descendants.*

| Closeness Centrality | | Betweenness Centrality | | Output domain | | Output proximity prestige | | Citation weight | | Supervising > 10 (abridged; number in parenthesis denotes ranking in full list) | | Examining > 10 | |
|---|---|---|---|---|---|---|---|---|---|---|---|---|---|
| AU_16069 | 0.12 | AU_16069 | 0.06 | TempID_219 | 747 | AU_16069 | 0.03 | TempID_219 | 0.06 | AU_16114 (2) | 29 | AU_16114 | 26 |
| AU_16148 | 0.12 | AU_16148 | 0.03 | AU_16069 | 740 | TempID_219 | 0.02 | AU_16069 | 0.06 | AU_16017 (3) | 27 | AU_16017 | 20 |
| AU_13024 | 0.12 | Merge_09577 | 0.03 | Merge_11021 | 325 | AU_16091 | 0.01 | Merge_11021 | 0.03 | AU_16069 (5) | 23 | Merge_09577 | 16 |
| AU_16091 | 0.12 | AU_16005 | 0.02 | TempID_256 | 323 | AU_16075 | 0.01 | AU_16068 | 0.03 | AU_16019 (8) | 20 | AU_16375 | 15 |
| AU_16088 | 0.12 | AU_16091 | 0.02 | AU_16068 | 322 | AU_16088 | 0.01 | TempID_256 | 0.03 | AU_16375 (9) | 20 | AU_13219 | 14 |
| AU_16114 | 0.11 | AU_16088 | 0.02 | AU_16091 | 317 | AU_16017 | 0.01 | AU_16091 | 0.08 | Merge_09577 (10) | 20 | AU_16019 | 12 |
| Merge_09577 | 0.11 | AU_16017 | 0.02 | AU_16075 | 306 | Merge_11021 | 0.01 | AU_16075 | 0.02 | AU_16024 (11) | 19 | AU_16097 | 11 |
| AU_13242 | 0.11 | AU_13024 | 0.02 | AU_16088 | 268 | AU_16068 | 0.01 | TempID_241 | 0.02 | AU_16091 (14) | 17 | AU_16019 | 11 |
| AU_16017 | 0.11 | AU_13197 | 0.02 | Merge_03510 | 188 | AU_13024 | 0.01 | AU_16088 | 0.02 | AU_13219 (16) | 16 | AU_16148 | 10 |
| AU_16084 | 0.11 | AU_16024 | 0.02 | | | TempID_241 | 0.01 | Merge_03510 | 0.01 | AU_16088 (21=) | 14 | AU_16091 | 10 |
| | | | | | | | | | | AU_16096 (21=) | 14 | Merge_05008 | 10 |
| | | | | | | | | | | Merge_04834 (24) | 13 | AU_16069 | 10 |
| | | | | | | | | | | AU_13024 (25) | 12 | Merge_04834 | 10 |
| | | | | | | | | | | AU_16148 (28=) | 11 | | |

**Occurances:** 7 | 6 | 5 | 4 | 3 | 2 | 1

*Table 5: Ranking of academics, with node ID's associated with names, from Table 4 with the same colour coding. We flag were an academic has only appeared in centrality measures.*

| Node id | Name | Appears in # measures | Appears only in centrality measures |
|---|---|---|---|
| AU_16069 | Washburn, Sherwood | 7 | |
| AU_16091 | Tuttle, Russell | 7 | |
| AU_16088 | Devore, Irven | 6 | |
| AU_16017 | Jolly, Clifford | 5 | |
| AU_13024 | Glander, Kenneth | 4 | |
| AU_16148 | Fleagle, John | 4 | |
| Merge_09577 | Rodman, Peter | 4 | |
| AU_16068 | McGrew, William | 3 | x |
| AU_16075 | Simons, Elwyn | 3 | x |
| AU_16114 | Sussman, Robert | 3 | |
| Merge_11021 | Srebnik, Herbert | 3 | x |
| TempID_219 | Hooton, Earnest | 3 | x |
| TempID_241 | Jepsen, Lowell | 3 | x |
| AU_13219 | Wright, Patricia | 2 | |
| AU_16019 | Wrangham, Richard | 2 | |
| AU_16024 | Richard, Alison | 2 | |
| AU_16096 | Poirier, Eugene | 2 | |
| AU_16097 | Bramblett, Claud | 2 | |
| AU_16375 | Dolhinow, Phyllis | 2 | |
| Merge_03510 | Dunbar, Robin | 2 | |
| Merge_04834 | Holekamp, Kay | 2 | |
| TempID_256 | Keith, Arthur | 2 | x |

In Table 4: Lists the top ten individuals for five different network measures (and those supervising/examining >10 theses). We then removed those identified by only one measure then look at the individuals who are prominent by several different measures. Table 5 combines a number of different network statistics that we deem to be important in assessing the importance of an individual to the social community. Twenty-two individuals are identified by at least two of the different measures considered and we discuss these below.

We note that the following six individuals are identified by centrality measures but do not appear in the lists of examinations or supervisors: Elwyn Simons, Herbert Srebnik, Earnest Hooton, Lowell Jepson, Arthur Keith, William McGrew.  Among these is Earnest Hooton (d 1954), Washburn's supervisor. One interpretation of Hooton's position is that his centrality is an effect of that of his student, he has, as it were, acquired some of the prestige of his illustrious student. Arthur Keith (d 1955) is the other older figure in our list, and a surprising one: he is usually regarded as a human paleontologist although he did study primate skeletons. The two measures in which he scores well are both those which, in effect, reflect the prominence/success of descendants ('output domain' and 'citation weights'. For the latter measure, readers should remember that these are not actual publication citations, rather we are using a measure first developed to analyse networks of citation patterns).

Conversely, we also note that the list includes Kay Holekamp, a senior zoologist who is not a primatologist by specialization. She enters our list as a prominent supervisor and examiner of primatologists.

Herbert Srebnik, Professor of Anatomy at Berkeley is of note since he is not a prolific examiner or supervisor nor has he been widely recognised by the award of medals etc. However, he has clearly played an important role as reflected in his pivotal position in the networks of supervision and examining.

### 3.  Esteem propagation and ranking

In this section, we extend the ideas of the previous section to explore our genealogy network in the context of ranking theory.  Here we wish to improve our "computation intuition" by adding features to our graph that allow us to not only assess the relative esteem of individuals but see how esteem "flows" across a network through interactions provided by PhD examinations and doctoral supervisions.

As stated at the beginning of this paper, esteem flows through an academic network. A reader's first impression of someone's curriculum vitae is influenced by the university that they attended, their doctoral advisor, examiners and a number of other factors. If we consider each participant in the network to be a node with a "score" defined by some objective assessment of their vitae and links between nodes provided by an academic hierarchy such as PhD mentorship or PhD examination (or a combination of these two) then we define a "system of esteem". If we wanted to understand the relative esteem of one person with respect to another in the network it is important to understand the context surrounding each person and one can use network analytic approaches to quantify the relative esteem of any two people on a network that is "strongly connected". (We won't define technical terminology here since it is not needed to follow our arguments, but instead refer the reader to [1,20].) Hence, when we talk about esteem flowing around a network, the concept is clear, each connection shared between people allows a little "glory" to be reflected in each direction. It is important to understand that when we talk about flow, we are not talking about a dynamical quantity on the network. Rather we are describing the rebalancing of the esteem scores at each point in the network when new information is added to the network. New information in this context can come from three types of event: i) the addition of a new edge – where a new qualification event takes place and links nodes that already exist in the network; ii) the addition of a new node as a result of activity i); and iii) the exogenous addition of new esteem – for example, a new prize or honour conferred a participant in the network. Assessing this network to calculate the relative esteem of the participants can be classed as a ranking problem. This is a well-studied problem that has been seen in many contexts from Google's PageRank algorithm to modern content recommendation engines and even College Football rankings [1, 20].

Typically, network approaches require a way to assign scores to each node in the graph to create a "preference matrix". This preference matrix is then used to create a ranking that takes into account not only the strength of each individual node individually, but the strength of the nodes that have relationships with that node, as well as second, third, and eventually $n$th level relationships. The work of Keener [1] shows that the construction of the preference matrix in a correctly-normalised manner preserves the probabilistic interpretation of the network and means that network statistics such as betweenness centrality and closeness centrality continue to be valid concepts.

While the network that we've studied is directed in time, in the sense that teachers are the progenitors of their students, the flow of esteem does not only travel "forward" in the graph but backward in time as well. For example, from our previous discussion, it is clear that having a PhD advisor or examiner who is the recipient of a Nobel Prize can be good for the student or examinee. However, the advisor to a student who goes on to win a Nobel Prize also benefits from an enhanced reputation. Hence, a network of academic relationships with different esteem factors associated with different individuals gives rise to a constantly rebalancing and shifting system of esteem.

An excellent example of PhD advisor who never won a Nobel Prize himself is the German physicist Arnold Sommerfeld who, as a PhD advisor produced Werner Heisenberg, Wolfgang Pauli, Peter Debye and Hans Bethe, all of whom won Nobel Prizes in Physics or Chemistry.  These were not the only famous names that he supervised (although not necessarily at doctoral level) he worked with Linus Pauling and in more mathematical matters with Ludwig Hopf, Rudolf Peierls and Wilhelm Lenz, all of whom made significant innovations in their chosen fields that are in use regularly by modern-day practitioners.  Sommerfeld himself is well known to field theorists for the Bohr-Sommerfeld quantisation condition and a number of other techniques.  He won the Max-Planck medal, the Lorentz Medal and become a Fellow of the Royal Society – all significant measures of esteem in their own right but not perhaps with the lustre of a Nobel Prize.  Yet Sommerfeld's reputation is certainly enhanced by having been associated with some of the great names in quantum physics and having supervised more Nobel Prizes winners than anyone else (with the possibly exception of the English physicist JJ Thompson).  Conversely, Richard Feynman dominated theoretical physics in the 1950s and 1960s winning a Nobel Prize for his contribution and is occasionally described as the "smartest man alive" in that period.  Yet, none of his students won a Nobel Prize.

In formulating our ranking we need to associate scores with each researcher.  In the ranking created by Keener [1] for football teams, this score could be derived from the number of goals achieved by each side.  Hence, in the Keener case, each game or interaction relates a pair of teams and these scores are summed over the season.  In our academic case, the picture is more static.  A scoring method needs to be defined such that when an honour is conferred on an individual, the node weight is adjusted to take this into account.  This is where a lot of subjectivity is involved in any ranking scheme. We see precisely this approach applied to university rankings, where different data and different weighting schemes lead to different outcomes.  Here we are attempting to make personal rankings more reflective of the context of the person (i.e. taking account of the interactions that we can derive from the relationships that people have in the form of qualification events, which are the analogue in this model of the football game interactions).  The effect of using the approach in [1] to rank the nodes in the network is that the final ranking is not simply the sum of the weights associated with the awards that each person has but rather it takes into account the network effects of being located in the network in proximity to an award winner.  This makes sense since, if a prize winner chooses a doctoral student, then they are likely to have had a good field to choose from and the chances that the student may be good is higher.  There is also a reinforcing effect where that student, being supervised by a prize winner, is more likely to be afforded opportunities than other students.  The network ranking approach associated with [1] models this type of reinforcement in a natural way.  The ranking scheme is also quite robust, which is to say that the model deals with partial information quite well. Typically, if scores can be assigned to the square root of the number of nodes in the network, a workable ranking can still be derived with some confidence.

In order to specify the elements of the preference matrix described above, we need to suggest a scoring methodology. As noted earlier, each node corresponds to either a supervisor, an examiner or a degree candidate.  The weight that we assign to each node in the network should describe their achievements or the esteem that has been conferred on the individual so far. Hence, our sketch formula for the weight of each node/researcher might be:

$$Weight = a \sum Nobel\ Prizes + b \sum Elected\ Fellowships + c \sum Awards$$

$$+ d \sum Honorary\ Degrees + e \sum Editorial\ board\ memberships$$

$$+ f \sum Grant\ committee\ memberships + \cdots$$

where the sum in each case is over all the awards won in each category by the person in their career so far, and each coefficient (a,b,c,…) represents a universal weight associated the type of award in the sum. Ideally, these universal weight factors are determined by surveying a large population of practitioners so that the relative strength of the awards are well understood. In practice, and in many cases where such methods are used in university rankings, these weight factors are either arbitrarily decided, hidden from those being ranked, or both. We normalise these weights (a, b, c, …) by summing across the nodes in the graph and dividing out. We note that modifying the weight function above by removing the co-efficients and summing reduces to the sum of awards given. Hence, the sum of the weights is a quantity that is something like the total "esteem" given. These concepts are analogous to particle number and energy preservation in a closed physical system. The weight function can be directly considered to be the "bare" esteem associated with an individual, however, it does not directly represent the network effect of being connected to other nodes in the graph. The output of the ranking approach described here would be a "ranking vector" that gives a relative score to each node in the graph. This score can be thought of as the relative network esteem or perceived esteem of a participant in the network –an equivalent to PageRank for Academic Esteem!

There are many more element types than we have listed in this short example that may be important to consider, for example these may include many well-established measures of esteem such as honorary chairs, named chairs, prestigious committee memberships, government advisory positions, through to more modern esteem measures such as having a Wikipedia page. In each case, there are subtle interplays between different elements, which need to be understood to calculate an overall score. For instance, the example above contains no interaction terms between awards. An example where it may be necessary to consider such interactions is the winning of a Fellowship of the National Academy of Science in the US. When an American researcher receives the Nobel Prize in their field then they will almost without question by invited to become a member of the National Academy of Science. Is it appropriate that they should get a "booster" to their esteem score by this or is it entirely right and the halo effect of a Nobel Prize should propagate unseen in our equation?

The prizes and elected fellowships as well as other awards and measures of prestige that we suggest above have a constancy or a quality that we contend is approximately constant in time. Hence, we borrow the concept of "standard candle" from astronomy. In astronomy, standard candles are types of celestial object that burn with a luminosity that is well understood and hence they can be used to calculate their distance. We posit, for an example, that a Nobel Prize always takes about the same level of academic achievement and command of a field relative to others and that, as a result of winning such a prize, the esteem of the recipient will manifest in approximately the same way at any point in time. Hence, by using the Esteem measure that we suggest above, we can compare the relative esteem of researchers across generations.

Even for a small, tight-knit community such as primatology, not everyone is connected in a single graph. There are colleagues who work on distinct topics and who form disconnected sub-groups. In Table 1, we summarize the sizes of the largest connected components of the graphs that we have worked with in this paper. The large connected components contain 65-70% of the nodes in the data. The network ranking approach that we have suggested here does not have the capacity to create a full relative ranking of all components if they are not part of the same connected piece. In exploring the problem further, we would need to work with the largest connected component for the chosen ranking method to be applicable.

Collating information about honours proved difficult. This was partly due to the difficulties of handling heterogeneous data. In particular, we encountered several questions as we looked deeper into this line of exploration: Which awards do we decide to be significant? Does the significance of awards change with time? Are records of award that are publicly available sufficiently complete? Other issues are also raised: How does one assess the relative merit of being elected to membership of Sigma XI or the British Academy over having discovered a new species? On the basis that success breeds success or that consistent success is reflected over a career by the accrual of awards and prizes we made a crude count of identifiable honours.

When counting the measures of esteem awarded to a key group of 22 individuals identified by the network analysis in Tables 4 and 5 there is no clear correlation. Only Russell Tuttle, scores highly on both, and there is no statistically significant correlation (see Table 6). Table 6 includes counts of many honours including: Fellowship American Association for the Advancement of Science, Fellowship of the American Academy of Arts and Sciences, the Alban-Heiser Award for Conservation of the Houston Zoological Society (for a full list please see ancillary material).

*Table 6: List of academics showing correlation between number of real-world honours received (Honour Count) and significant position in academic network analysis (Top 20 nodes by network centrality measure).*

| Name | Honour Count | Network Measures Count |
|------|:---:|:---:|
| Dolhinow, Phyllis | 1 | 2 |
| Poirier, Eugene | 1 | 2 |
| Rodman, Peter | 1 | 4 |
| Srebnik, Herbert | 1 | 3 |
| Bramblett, Claud | 2 | 2 |
| Sussman, Robert | 2 | 3 |
| Jolly, Clifford | 3 | 5 |
| Holekamp, Kay | 5 | 2 |
| Jepsen, Lowell | 5 | 3 |
| Keith, Arthur | 7 | 2 |
| Washburn, Sherwood | 7 | 7 |
| Glander, Kenneth | 9 | 4 |
| Devore, Irven | 13 | 6 |
| Dunbar, Robin | 16 | 2 |
| Hooton, Earnest | 16 | 3 |
| Simons, Elwyn | 16 | 3 |
| Fleagle, John | 19 | 4 |
| Tuttle, Russell | 21 | 7 |
| Wrangham, Richard | 25 | 2 |
| Richard, Alison | 26 | 2 |
| McGrew, William | 27 | 3 |
| Wright, Patricia | 37 | 2 |

## 4. Discussion and future directions for research\

What this exercise has shown is that combining supervisory and examiners data is powerful. It enables us to identify individuals who are judged on the basis of their publications and other activity (including doctoral supervision and examining) to be of merit. In the UK context this means that metrics (or 'indicators' see Ref. 22) based on supervisory and external examination activity might be of interest to HEFCE in their regular Research Evaluation Framework exercises.

In the long term we hope that this pilot case will encourage the keepers of the metadata, the directors of DART (Digital Access to Research Theses), Proquest theses database and so on, to include examination data with agreed metadata standards so that this sort of analysis can be undertaken across all academic disciplines and evaluated across time.

There are many potential directions for new research projects based on the ideas in this paper and we only indicate a few here. It is interesting to solidify the model that we have sketched by surveying broadly to find appropriate weightings for each of the esteem factors in the weight equation.  Extending the model so that field dependence is taken into account with some appropriate normalisation is important.

A further fruitful area for future enquiry is the study of the relative weights for supervisors and examiners in the genealogy.  Can a generalised rule be developed for the relative weighting or should the weighting be dependent on the relative perceived strengths of the examiner and supervisor nodes?

It may also be interesting to study the "horizon of last influence".  That is, the longest time back that any "significant" change can be made to the perceived esteem of a colleague – a clear definition of "significance" needs to be defined for this to be possible.  We postulate that this is field independent since the effect of a supervisor's supervisor on a student may be minimal and a 3$^{rd}$ degree relationship is almost certainly even less.  However, particular fields may have especially close networks, particularly short generation times or the effects of major prizes such as the Nobel prize may have a disproportionate effect on the structure of those fields in which they exist.

Another direction is considering the use of community detection and pairing this with ranking to look at the distribution of awards on a naturally emergent subject categorisation.  It is clear that some research areas profit from more awards than others, hence, combining the work of Evans and Lambiotte [9,12], and Hutchins et al. [18, 21] suggests interesting possibilities for clustering and local rankings within academic "families".

To summarise, we have demonstrated that intellectual genealogies can be enhanced by going beyond the supervisor – student relationship, and that to properly understand the lineages we need to consider bilateral kinship structures.  The important figures in the networks are those who are recognized as high status through the award of prizes, medals and prestigious fellowships.  These prizes form the basis of our perception of who is most influential in academic networks.  However, prestige/esteem is a more complicated quantity and proximity to those who hold prestigious awards in the academic genealogy can be a signal for unseen influence or future success depending on career stage.

**Acknowledgments**

We thank the following for helping us collect and collate the data: Andrew Wheale, Bahar Tuncgenc and Daniel Mullins.  Proquest and the various intellectual genealogy projects as well as Kelley and Sussman and many other individuals. Network analysis and visualisation was helped by Martin Hadley at University of Oxford IT Services and by Frank Hangler at Plot and Scatter.

**Funding Statement**

Digital Science provided support in the form of salary for DWH and provided funding to support data analysis and travel to meetings but did not have any additional role in the study design, data collection and analysis, decision to publish, or preparation of the manuscript. This work, conducted in 2015 and 2016, has informed Digital Science's innovation work.  At the date of submission no product has been brought to market that makes use of this research.